\documentclass{ws-p8-50x6-00}
\usepackage{wrapfig}  

\def\la{\mathrel{\mathchoice {\vcenter{\offinterlineskip\halign{\hfil
$\displaystyle##$\hfil\cr<\cr\sim\cr}}}
{\vcenter{\offinterlineskip\halign{\hfil$\textstyle##$\hfil\cr<\cr\sim\cr}}}
{\vcenter{\offinterlineskip\halign{\hfil$\scriptstyle##$\hfil\cr<\cr\sim\cr}}}
{\vcenter{\offinterlineskip\halign{\hfil$\scriptscriptstyle##$\hfil\cr<\cr
\sim\cr}}}}}

\begin{document}

\title{Cosmic Rays in the Energy Range of the Knee\\
-- Recent Results from KASCADE --}

\author{K.-H.~Kampert$^{1,2}${\footnote{E-mail: kampert@ik1.fzk.de}},
T.~Antoni$^{2}$,
W.D.~Apel$^{2}$, 
F.~Badea$^{3}$, 
K.~Bekk$^{2}$, 
K.~Bernl\"{o}hr$^{2}$,
H.~Bl\"{u}mer$^{2,1}$,
E.~Bollmann$^{2}$, 
H.~Bozdog$^{3}$, 
I.M.~Brancus$^{3}$, 
C.~B\"{u}ttner$^{2}$,
A.~Chilingarian$^{4}$,
K.~Daumiller$^{1}$,
P.~Doll$^{2}$,
J.~Engler$^{2}$,
F.~Fe{\ss}ler$^{2}$,
H.J.~Gils$^{2}$,
R.~Glasstetter$^{1}$,
R.~Haeusler$^{2}$,
A.~Haungs$^{2}$,
D.~Heck$^{2}$,
T.~Holst$^{2}$,
J.R.~H\"{o}randel$^{1}$,
A.~Iwan$^{5}$,
J.~Kempa$^{5}$,
H.O.~Klages$^{2}$,
J.~Knapp$^{1}$,
G.~Maier$^{2}$,
H.-J.~Mathes$^{2}$,
H.J.~Mayer$^{2}$,
J.~Milke$^{2}$,
M.~M\"{u}ller$^{2}$,
J.~Oehlschl\"{a}ger$^{2}$,
M.~Petcu$^{3}$,
H.~Rebel$^{2}$,
M.~Risse$^{2}$,
M.~Roth$^{2}$,
G.~Schatz$^{2}$,
H.~Schieler$^{2}$,
J.~Scholz$^{2}$,
T.~Thouw$^{2}$,
H.~Ulrich$^{1}$,
B.~Vulpescu$^{3}$,
J.H.~Weber$^{1}$,
J.~Wentz$^{2}$,
J.~Wochele$^{2}$,
J.~Zabierowski$^{6}$,
S.~Zagromski$^{2}$
}

\address{$^{1}$ Institut f\"{u}r Exp. Kernphysik, Universit\"{a}t 
Karlsruhe (TH), (Germany)\\
$^{2}$ Institut f\"{u}r Kernphysik, Forschungszentrum Karlsruhe,
(Germany)\\
$^{3}$ Institute of Physics and Nuclear Engineering, Bucharest,
(Romania)\\
$^{4}$ Cosmic Ray Division, Yerevan Physics Institute, (Armenia)\\
$^{5}$ Department of Experimental Physics, University of Lodz,
(Poland)\\
$^{6}$ Institute for Nuclear Studies, Lodz, (Poland)
}

\maketitle

\abstracts{A brief motivation for studying cosmic rays at
energies $10^{14} \la E \la 10^{17}$ eV is given.  Besides
astrophysical interests in identifying and understanding their
sources, there are also particle physics aspects related to their
transport properties in the galaxy or to their detection via
extensive air showers.  The KASCADE air shower experiment taking
data at Forschungszentrum Karlsruhe (Germany) provides important
information to both of these topics of particle-astrophysics
research.  A target of particular interest is the so-called
``knee'' in the cosmic ray energy spectrum at $E_{\rm k} \approx
4 \cdot 10^{15}$ eV. Recent results adding knowledge to an
understanding its origin will be presented as well as results on
tests of high-energy hadronic interaction models required for air
shower simulations.}

\section{Introduction}

The origin and acceleration mechanism of ultra-high energy cosmic
rays have been subject to debate for several decades.  Mainly for
reasons of the power required to maintain the observed cosmic ray
energy density of $\varepsilon_{\rm cr} \approx 1$ eV/cm$^{3}$
the dominant acceleration sites are generally believed to be
supernova remnants (SNR).  Charged particles mainly originating
from the surrounding interstellar medium of the pre-supernova
star may get trapped at the highly supersonic shock wave
generated by the SN explosion.  Repeatedly reflections on both
sides of the shock front lead to an acceleration by the so-called
`first order Fermi mechanism'.  Naturally, this leads to a power
law spectrum $dj/dE \propto E^{-\gamma}$ as is observed
experimentally.  Simple dimensional estimates show that this
process is limited to $E_{\rm max} \la Z \times (\rho \times B)$,
with $Z$ being the atomic number of the cosmic ray (CR) isotope
and $\rho$, $B$ the size and magnetic field strength of the
acceleration region.  A more detailed examination of the
astrophysical parameters suggests an upper limit of acceleration
of $E_{\rm max} \approx Z\times
10^{15}$\,eV\cite{drury94b,berezhko99}.  Curiously, the CR
spectrum steepens from $\gamma \simeq 2.75$ to $\simeq 3.1$ at $E
\simeq 4 \times 10^{15}$\,eV which is called the `knee'.  The
coincidence thus may indicate that the `knee' is related to the
upper limit of acceleration.

An alternative interpretation of the knee is that it may reflect
a change in the propagation of CRs from their sources to the
solar system.  Propagation effects are studied mostly by balloon
or space borne experiments measuring abundances of secondary to
primary CR isotopes at energies up to some GeV. A prominent
example is the B/C-ratio.  Analysing such data yields the `escape
time' of CRs from our galaxy which scales with energy according
to $\tau_{\rm esc} \propto E^{-\delta}$ with $\delta \approx
0.6$.  A simple extrapolation of $\tau_{\rm esc}$ to higher
energies breaks down at $E \approx 3\cdot10^{15}$ eV, because
$c\tau_{\rm esc} \sim 300$ pc which is the thickness of the
galactic disk\cite{gaisser00a}.  The value corresponds to just
one crossing of the disk and would give rise to significant
anisotropies with respect to the galactic plane when approaching
this value.  Similarly as to the process of acceleration at SNR
shocks, the process of galactic containment is closely related to
magnetic field confinement, i.e.\ in addition to anisotropies one
again expects $E_{\rm max}^{\rm gal} \propto Z$.

A picture related to both of these interpretations has been
proposed by Erlykin and Wolfendale\cite{erlykin97a}.  They
consider the knee as a superposition of a weakly energy dependent
galactic modulation with additional prominent structures in the
flux spectrum caused by a single near-by object.  This so-called
'single source model' assumes that a shock wave of a recent
nearby supernova which exploded some 10,000 years ago at a
distance of a few hundred parsecs currently propagates, or has
recently propagated through the solar system causing distinct
peaks of elemental groups in the energy spectrum.  However, there
is some controversy whether or not the statistics of presently
existing data gives support to the model.

Recently, a more particle physics motivated picture of explaining
the knee was put forward by Wigmans\cite{wigmans00}.  He
suggested that inverse beta decay of protons with massive relic
neutrinos according to $p + \bar{\nu_{e}} \to n + e^{+}$ could
destroy protons.  Simple kinematics shows that this channel is
open for $E_{p} > 1.7\cdot10^{15} {\rm eV}/m_{\nu} {\rm (eV)}$. 
Thus, the knee energy $E_{\rm knee} \simeq 4$ PeV would
correspond to an electron neutrino mass of $\sim$ 0.4 eV, a value
presently not excluded by any other observation or experiment. 
However, `eating' sufficiently large amounts of protons by such a
process requires extraordinary high local densities of relic
neutrinos, even if possible gravitational trapping is considered.

It has also been suggested that the knee is not a property of the
primary energy spectrum itself, but rather may be caused by
changing high-energy interactions in the earth
atmosphere\cite{nikolsky95}.  Producing a new type of a heavy
particle in the first interactions escaping unseen by air shower
experiments could, in principle, mimic a break in the spectrum. 
From the particle physics point of view this is not completely
ruled out as the centre-of-mass energy available at the knee is
above Tevatron energies.  Different from the astrophysical
interpretations of the knee, one would now expect to see the knee
energy of different primary particles being displaced by their
mass number $A$ rather than by their charge $Z$.  This is
understood from the nuclear reaction mechanism being governed by
the energy per nucleon $E/A$ of the incident particle.

A key observable for understanding the origin of the knee and
distinguishing the SN acceleration model from the other proposed
mechanisms, is thus given by CR energy spectra of different
elemental groups or, if measured more inclusively, by the mass
composition of CRs measured across the knee (see e.g.\
Ref.\cite{berezhko99}).  This is to be complemented by
measurements of anisotropies in the arrival directions of CRs. 
Finally, experiments should also aim at testing hadronic
interaction models by means of air shower data.

At present, beyond the knee little is known about CRs other than
their all-particle energy spectrum\cite{watson98}.  The low flux
of particles ($\sim 1$ m$^{-2}\cdot$year$^{-1}$ above the knee)
puts strong demands on the collection power of the experiments,
such as can at present only be achieved by extended air shower
(EAS) arrays at ground level.  The development of an EAS is a
process of truly multi-particle dynamics with frequent
high-energy hadronic and electromagnetic interactions taking
place.  At primary energies of $10^{15}$ eV about $10^{6}$
particles reach sea-level.  These are mostly photons and
electrons plus some fraction of high-energy muons and hadrons,
spread out over an area of several hectares.  Sampling detector
systems with typical coverages of less than one percent can thus
be used for registration of such EAS. However, this indirect
method of detection bears a number of serious difficulties in the
interpretation of the data and requires detailed modelling of the
air shower development (see talk given by
D.~Heck\cite{heck-ismd}) and detector responses.  Particularly,
systematic effects caused by the employed high-energy interaction
models and by inevitable EAS fluctuations need to be considered. 
The basic concept of the KASCADE experiment is to measure a large
number of EAS parameters in each individual event in order to
verify the consistency of EAS simulations and to determine the
energy and mass of the primary particles on a reliable basis.  It
is hoped that the new data will substantially improve our
knowledge of the origin of high-energy cosmic rays.

\section{KASCADE Experiment}

\begin{figure}[t]
\epsfxsize=\textwidth
\epsfbox{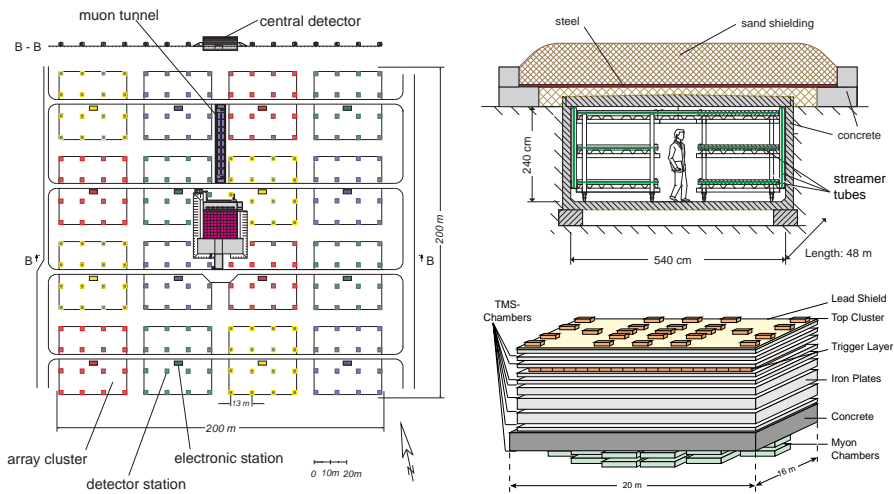}
\caption[]{Schematic layout of the KASCADE experiment (left), with 
its streamer tube tracking system (top right) and central 
detector (bottom right)\cite{kascade-97c}.}
\label{fig:kascade}
\end{figure}

KASCADE (\underline{Ka}rlsruhe \underline{S}hower \underline{C}ore and 
\underline{A}rray \underline{De}tector) is located at the laboratory 
site of Forschungszentrum Karlsruhe, Germany (at $8^{\circ}$ E, 
$49^{\circ}$ N, 110 m a.s.l.).  In brief, it consists of three major 
components (see Fig.\,\ref{fig:kascade});

\begin{enumerate}
    \item A scintillator array comprising 252 detector stations of 
    electron and muon counters arranged on a grid of $200 \times 200$ 
    m$^{2}$ and providing in total about 500 m$^2$ of $e/\gamma$- and 
    620 m$^{2}$ of $\mu$-detector coverage. The detection thresholds
    for vertical incidence are $E_{e} > 5$ MeV and $E_{\mu} > 
    230$ MeV.

    \item A central detector system (320 m$^{2}$) consisting of a
    highly-segmented hadronic calorimeter read out by 40,000
    channels of warm liquid ionization chambers, a layer of
    scintillation counters above the shielding, a trigger
    plane of scintillation counters in the third layer and, at 
    the very bottom, 2 layers of positional sensitive MWPC's,
    and a streamer tube layer with pad read-out
    for muon tracking at $E_{\mu} \ge 2.4$ GeV.

    \item A $48 \times 5.4$ m$^{2}$ tunnel housing three
    horizontal and a two vertical layers of positional sensitive
    limited streamer tubes for muon tracking at $E_{\mu} \ge 0.8$
    GeV.

\end{enumerate}

More details about the experiment can be found in Refs.\
\cite{kascade-97c,kascade-90}.  First correlated data have been
taken with some parts of the experiment in 1996 and with its full
set-up since 2000.  At present, more than 300 Mio.\ events have
been collected in a very stable mode and with a trigger threshold
of the array corresponding to $E \sim 4 \cdot 10^{14}$ eV.

\section{Tests of High-Energy Hadronic Interaction Models}

\begin{figure}[t]
\epsfxsize=8.5cm
\centerline{\epsfbox{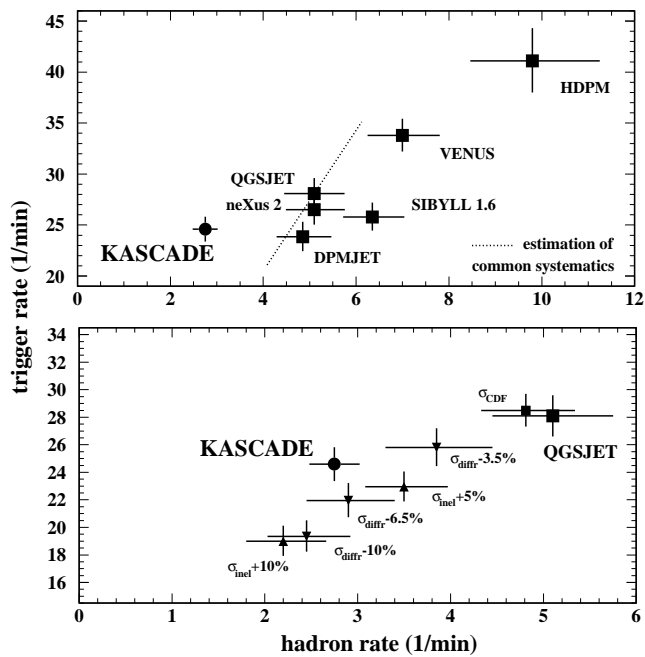}}\vspace*{-3mm}
\caption[]{Trigger rate vs hadron rate in KASCADE. The top panel
compares different hadronic interaction models to the
experimental data.  The systematic uncertainty, mostly given by
the absolute flux uncertainty of the direct experiments, is
indicated by the dotted line.  The lower panel shows results
obtained from the QGSJET\cite{qgsjet} model with modified
inelastic cross section and diffraction
dissociation\cite{risse99,kascade-01b}.}
\label{fig:trig-rates}
\end{figure}

The observation of EAS provides an opportunity to study global
properties of hadronic interactions in an energy range not
accessible to man-made accelerators.  For example, the
centre-of-mass energy at the Tevatron collider corresponds to a
fixed target energy in the nucleon-nucleon system of $E_{p}
\simeq 1.7 \cdot 10^{15}$\,eV. Even more importantly, the
diffractive particle production dominating the energy flux in the
forward region and influencing the EAS development most strongly,
was studied experimentally only at comparatively low energies of
$\sqrt{s}\simeq 10$ GeV\cite{kaidalov79}.  Most of the beam
energy in present collider experiments, however, remains unseen. 
For example, the UA5 experiment could register up to 30\,\% of
the total collision energy at $\sqrt{s}=0.9$ TeV, while the CDF
detector registers only about 5\,\% at $\sqrt{s}=1.8$ TeV. Hence,
hadronic interaction models applied to higher energies and to
particle production in the forward region rely on extrapolations
and may cause systematic uncertainties in simulations of
EAS. Additional uncertainties arise from simulations of p-nucleus
and nucleus-nucleus collisions including a possible formation of
a quark-gluon plasma.  Again, such data are important for EAS
interpretations but have been studied only at low energies in the
past (SPS and ISR at CERN).  Only very recently, RHIC data at
$\sqrt{s} = 200$ GeV have become available and will be very
helpful in this respect.  Apart from different theoretical
extrapolations, an additional source of uncertainty for all
models originates from the experimentally deduced energy
dependent inelastic cross-sections\cite{block00}.  At Tevatron
energies, the proton-antiproton cross section is not known to
better than 5\,\% at best.  As will be shown below, these
uncertainties are amplified when predicting absolute hadron
fluxes at sea-level by means of EAS simulations.  Thus, one may
take advantage of this strong dependence and perform stringent
tests of models on the basis of EAS data.  The idea is to use data of
{\em absolute} CR fluxes up to several TeV of energy, as obtained
from balloon and satellite experiments at the top of the
atmosphere, and propagate these particles (taking into account
their energy distribution and chemical composition) through the
atmosphere employing the CORSIKA simulation
package\cite{corsika}.  At the level of the KASCADE experiment,
we then ask for triggers released by either high energy hadrons
($E_{h} > 90$ GeV) or a minimum number of 9 muons detected in
the central detector of KASCADE. These simulated inclusive
trigger rates, employing different hadronic interaction models, are
then compared to actual experimental
data\cite{risse99,kascade-01b}.  Figure \ref{fig:trig-rates}
shows the results.  None of the predictions agrees well with the
experimental data, particularly the hadron rates are
overestimated by up to a factor of 3.  Furthermore, there are
also large differences found between the models.  This
convincingly proves the sensitivity of the experimental
observable to details of the interaction models.

Earlier investigations of high-energy hadrons observed in the
shower core of the KASCADE calorimeter\cite{kascade-99c} lead to
the conclusion that QGSJET provides the best overall prescription
of EAS data.  This was confirmed also by an independent analysis
including data from different EAS experiments\cite{erlykin98d}. 
Therefore, several modifications were applied to the QGSJET model
in order to study their influence to the predicted trigger rates. 
Results are presented in the lower panel of
Fig.\,\ref{fig:trig-rates}.  Increasing the inelastic proton-air
cross section by 5 and 10\,\% reduces the predicted hadron rate
by approx.  27\,\% and 54\,\%, respectively.  Similar effects are
obtained by lowering the diffraction dissociation by up to 10\,\%
of the inelastic cross section.  Clearly, there is a strong
sensitivity to these parameters.  Since the total inelastic cross
section appears to be the better known quantity, we conclude that
the diffraction dissociation is overestimated in the simulated
hadron-nucleus interactions by about 5\,\% of the inelastic cross
section $\sigma_{\rm inel}$(p-air).  Further studies with new
models and with high energy hadrons observed in the core of EAS
are in progress and will, hopefully, help to further improve our
understanding of hadronic interaction processes in EAS. This will
be important also for interpretations of EAS data in terms of
primary energy and chemical composition.

\section{Energy Spectrum and Chemical Composition}

Extracting the primary energy and mass from EAS data is
not straightforward and a model must be adopted to relate the
experimental observables to properties of the primary
particle.  As shall be discussed below, the analysis is
complicated by large fluctuations of EAS observables and by the
fact, that virtually all of these observables are sensitive to
changes in the mass {\em and} energy of the primary particle.

Air shower measurements generally include particle measurements
at ground and are sometimes complemented by detection of
Cherenkov- and fluorescence light\cite{kampert01a}.  Primary
observables of KASCADE are lateral particle density
distributions, $\rho_{e,\mu,h}(r)$, and their integrals
$N_{e,\mu,h} = \int_{r_{1}}^{r_{2}} 2\pi r \rho_{e,\mu,h}(r) dr$,
yielding total (if extrapolations $r_{1} \to 0$ and $r_{2} \to
\infty$ are made) or truncated particle shower sizes.  Truncated
shower-sizes, $N_{e,\mu,h}^{\rm tr}$, have been introduced by the
KASCADE collaboration in order to avoid systematic uncertainties
resulting from extrapolations of $\rho_{e,\mu,h}(r)$ to large
distances not covered by the experiment.  Reconstruction
techniques of the individual particle densities, shower core
position, and shower direction have been discussed in
Ref.\,\cite{kascade-01a}.  As simulations predict and as it is
obvious from very basic considerations, the weighted sum of
particle shower sizes provides an estimate of the primary energy,
i.e. $E \simeq a \cdot \log N_{\mu}^{\rm tr} + b \cdot \log
N_{e}^{\rm tr}$, with $a$ and $b$ determined from simulations. 
On the other hand, the ratio $Y_{\rm ratio} =\log N_{\mu}^{\rm
tr} / \log N_{e}^{\rm tr}$ provides a
\begin{wrapfigure}[20]{l}{6cm}
\epsfxsize=6.3cm
\epsfbox{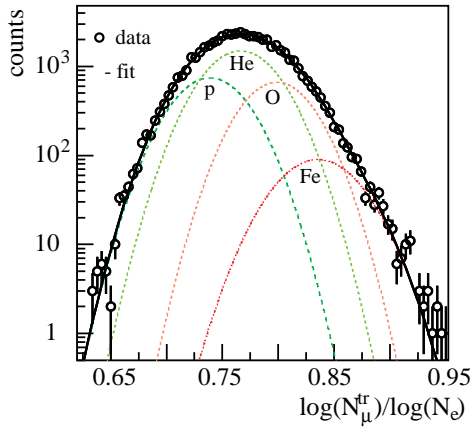}
\vspace*{-7mm}
\caption[]{Example of an event-by-event distribution of the
muon/electron ratio measured by KASCADE for a primary energy of
$E \simeq 3\cdot 10^{15}$ eV. The lines show results of CORSIKA
simulations employing the QGSJET model and are performed for
different primary particles.}
\label{fig:mu-e-ratio}
\end{wrapfigure} 
reliable estimate of the
primary mass\cite{kascade-99d}.  This is mostly because showers
of heavy primaries will, on average, develop higher in the
atmosphere so that the electromagnetic component suffers more
absorption in the atmosphere as compared to a shower induced by a
primary proton\cite{kampert01a}.  An example is presented in
Fig.\,\ref{fig:mu-e-ratio}.  Clearly, a superposition of
different particle types (represented by the lines) is required
to account for the experimental distribution.  Noteworthy, the
left and right hand tails of the experimental distribution are
nicely described by the proton and iron simulations,
respectively.  Performing such an analysis for different bins of
shower sizes yields an increasingly heavier composition for
energies above the knee\cite{kascade-99d}.

\begin{figure}[t]
\epsfxsize=0.7\textwidth
\centerline{\epsfbox{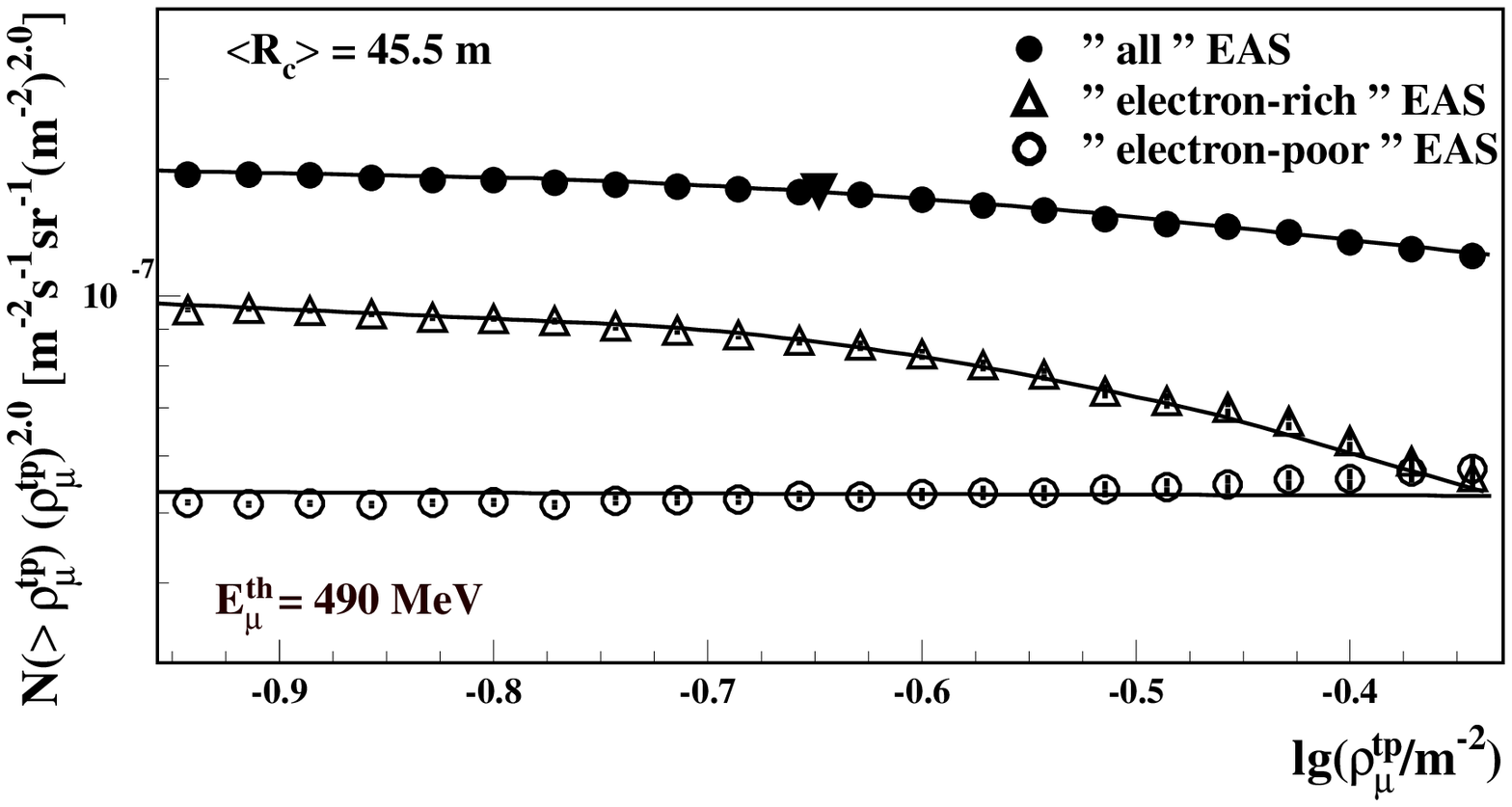}}
\caption[]{Distribution of muon densities $\rho_{\mu}$ ($E_{\mu}
> 490$ MeV) measured in the trigger plane at a distance of 45.5 m
from the shower axis.  The `all'-particle spectrum is shown as
well as spectra for `electron-rich' and `-poor' showers.  The
black triangle indicates the position of the
knee\cite{kascade-01c,haungs99}.}
\label{fig:mu-density}
%
\epsfxsize=0.65\textwidth
\centerline{\epsfbox{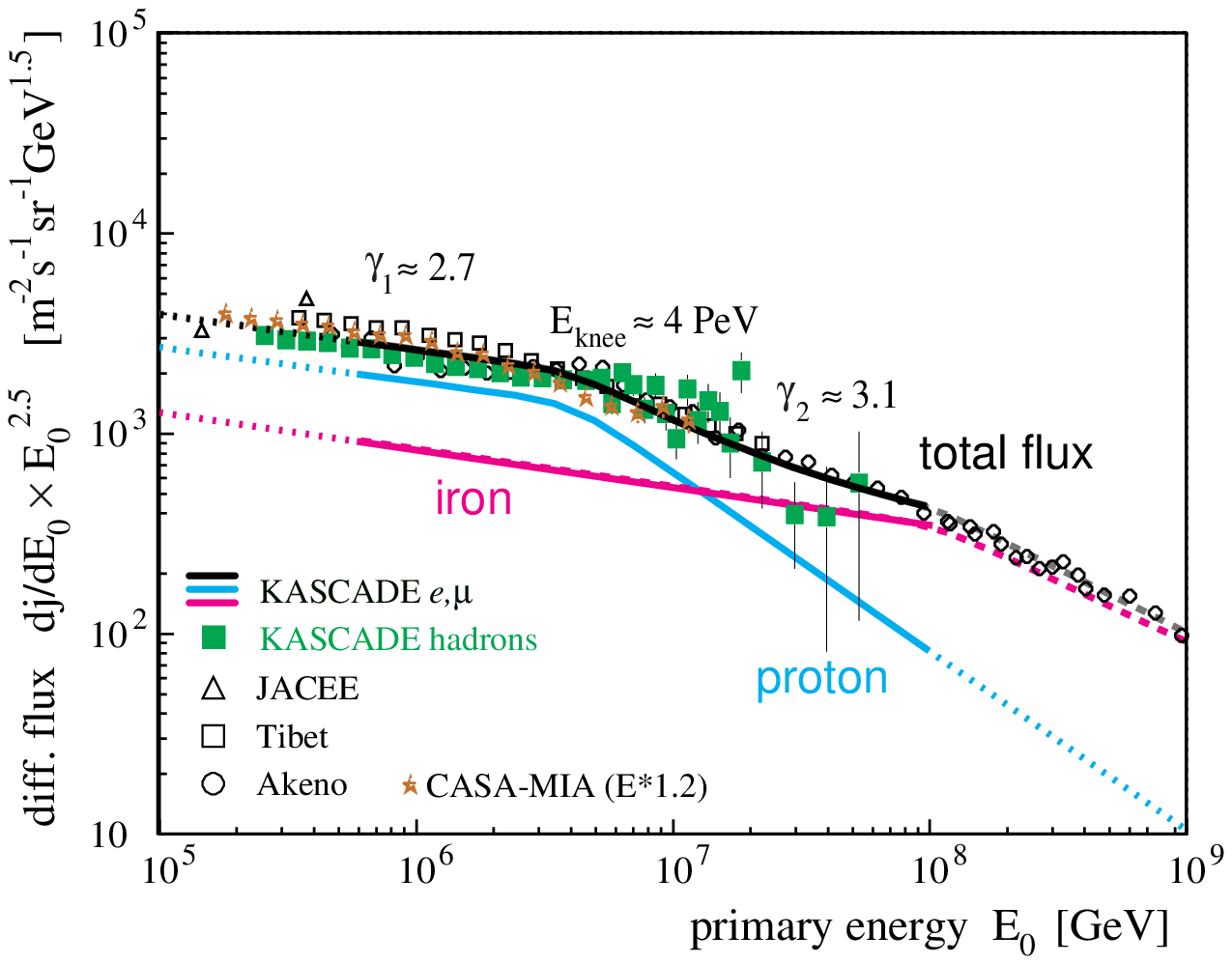}}
\caption[]{Primary energy spectrum as obtained from different
experiments.  The lines represent
KASCADE results based on a simultaneous fit to electron and muon
shower sizes assuming the all-particle spectrum to be a
superposition of a proton and iron
component\cite{glasstetter99}.}
\label{fig:e-spec}
\end{figure}

The separation of primary particle types by $Y_{\rm ratio}$ may
also be used to disentangle the primary flux spectrum into
`light' and `heavy' components.  An example is shown in
Fig.\,\ref{fig:mu-density}.  Here, the distribution of muon
densities measured in the trigger plane of the central detector
at a distance of 45.5\,m from the shower core is plotted for all
showers and for those where $Y_{\rm ratio} < 0.75$
(`electron-rich') and $Y_{\rm ratio} > 0.75$ (`electron-poor'). 
Interestingly, the knee structure is seen for unselected and
electron-rich showers but is absent for electron-poor
showers\cite{kascade-01c,haungs99}.  Realizing that the muon
density at given distance to the shower core is linked to the
primary energy, the figure proves in a very model independent way
that {\em the knee structure appears to be a feature caused by
the light component only}.  Again, we like to stress that
$\rho_{\mu}$ is directly measured by the experiment and that any
meaningful model will predict electron-rich and -poor showers for
light and heavy primaries, respectively.

This finding is confirmed by combined energy and composition
analyses of the shower size spectra of KASCADE, albeit then based
on quantitative Monte Carlo simulations.  A compilation of the
all-particle energy spectrum reconstructed from various
experiments is presented in figure \ref{fig:e-spec}.  The
agreement appears reasonable and deviations are mostly explained
by uncertainties in the energy scale by up to 25\,\%, e.g.\ CASA
MIA data\cite{glasmacher99a} were shifted upwards in energy by
20\,\% to yield a better agreement to the other data sets.  The
shift is likely to be explained by the interaction model SIBYLL
1.6\cite{sibyll16} employed by the authors of
Ref.\,\cite{glasmacher99a} but which has been proven to provide
only a poor description of the experimental
data\cite{kascade-99c}.  The lines correspond to a simultaneous
fit of the electron and muon size spectra of KASCADE, assuming
the all-particle spectrum to be described by a sum of proton and
iron primaries\cite{glasstetter99}.  Energy and mass dependent
fluctuations of the electron and muon shower size are taken into
account in this analysis and, in fact, turn out to be important
for any quantitative discussion.  Again, the knee is only
reconstructed for the light component and no indication of a
break is seen in the heavy component up to about $10^{17}$~eV.
Obviously, the different spectral shapes of the `light' and
`heavy' induced air showers will result in an increasingly
heavier composition above the knee, such as is found also by
other experiments.  The analysis is an ongoing process with
improved statistics and reconstruction techniques making use of
the multi-detector capabilities of KASCADE.\cite{kascade-01d}

\section{Conclusions and Outlook}

KASCADE data provide mounting evidence for the knee being caused
by the light particle component only.  Up to an energy of $\sim
10^{17}$ eV, which is the upper limit of the present KASCADE
experiment, no indication of a similar structure is found in the
iron-like distribution.  However, comparison to the all-particle
spectrum deduced from other experiments, suggests the necessity
of a similar break in the heavy component just above $10^{17}$
eV. This important finding gives direct support to the picture of
acceleration or galactic confinement in magnetic fields.  To
discriminate the two pictures, high statistics data on anisotropy
will be required and are currently being analysed.  Furthermore,
an extension of the mass-selective measurements to higher
energies appears mandatory in order to test the hypothesis of a
`Fe-knee' in more detail.  This should go in parallel with further
tests of hadronic interaction models in order to improve the
reliability of the EAS data analysis.

The KASCADE and EAS-TOP\cite{aglietta89} collaborations have just
started a common effort to realise these goals already in the
near future.  The EAS-TOP scintillators are currently positioned
next to the KASCADE experiment at the site of Forschungszentrum
Karlsruhe providing a 12 times larger acceptance as compared to
the present set-up.  Generating a common trigger for the combined
experimental installations will allow taking advantage of the
full multi-detector capabilities including hadronic measurements
as well as muon counting, tracking, and timing capabilities.  The
new installation is foreseen to start taking data in spring 2002. 
It is expected that the upcoming KASCADE and future
KASCADE-GRANDE data will allow extending the tests of hadronic
interaction models to higher energies and to substantially
improve our understanding of the origin of the knee in particular
and about the origin and acceleration of cosmic rays in general.

\begin{frenchspacing}

\end{frenchspacing}

\end{document}